\documentstyle[aps,prl,multicol,psfig]{revtex}
\begin{document}
\title{Strong correlation, electron-phonon interaction and critical
fluctuations: isotope effect, pseudogap formation, and phase diagram of the  
cuprates}  
\author{C. Di Castro, M. Grilli, and S. Caprara}  
\address{Istituto Nazionale per la Fisica della Materia, UdR Roma 1 and  
Dipartimento di Fisica, Universit\`a di Roma ``La Sapienza'',\\  
Piazzale Aldo Moro 2, I-00185 Roma, Italy}  
\maketitle

\begin{abstract} 
Within the Hubbard-Holstein model with long-range Coulomb forces, we revisit 
the charge-ordering scenario for the superconducting cuprates and account for 
the presence or the absence of a formed stripe phase in different classes of 
cuprates. We also evaluate the mean-field and the fluctuation-corrected 
critical lines for charge ordering and we relate them with the various 
pseudogap crossover lines occurring in the cuprates and we discuss a mechanism
for their peculiar isotopic dependence. Considering the dynamical nature of 
the charge-ordering transition, we explain the spread of $T^*$ and of its 
isotopic shift, obtained with experimental probes with different 
characteristic time scales.
\end{abstract}  
  

\begin{multicols}{2}

\section{Introduction}
The phase diagram of the cuprates, like, e.g., 
${\rm La}_{2-x}{\rm  Sr}_x {\rm CuO}_4$ (LSCO), is characterized by three 
distinct regions which are reached by varying the doping $x$: (i) the 
overdoped region, where the system displays a markedly metallic character, 
which can be reasonably described within the standard Fermi-liquid picture; 
(ii) the region around optimal doping, where the superconducting critical 
temperature $T_c$ is maximum, and no other energy scale besides the 
temperature $T$ is seen in various experiments, so that the metallic phase 
above $T_c$ strongly deviates from the Fermi-liquid behavior; (iii) the 
underdoped region where, on the contrary, there is an abundance of energy 
scales, and an even stronger violation of the Fermi-liquid behavior. These 
are related to a doping-dependent temperature $T^0$, below which a 
suppression of the density of states at the Fermi level is revealed in static 
measurements and to a doping-dependent temperature $T^*$ below which  various 
experiments reveal a pseudogap $\Delta_p$. $T^*$ strongly depends on the 
characteristic time scale of the probe used to measure the pseudogap 
properties. Other energy scales of the underdoped phase are the 
superconducting gap $\Delta$ and the critical temperature $T_c$. 

The generic partition of the phase diagram into three qualitatively quite
distinct regions with markedly different behaviors and the absence of any 
energy scale but the temperature in the intermediate (optimally doped) region,
suggest the presence of a quantum critical point (QCP) near optimal doping 
\cite{BILLINGE,ZEIT,TALLONLORAM,BOEBINGER,PANAGOPOULOS,FENG}. If this is the 
case, the identification of the optimally doped region with the region where
quantum critical fluctuations are present naturally accounts for the peculiar 
non-Fermi-liquid behavior. Various proposals have been made to specify the 
broken symmetry phase to be associated with the underdoped region, ranging 
from a circulating-current phase \cite{VARMA}, to a change in symmetry of the 
superconducting order parameter \cite{SACHDEV}. The scenario that we have 
been proposing along the years \cite{ZEIT,RASS} is related to the occurrence
of an instability for charge ordering (CO), along a line $T_{CO}(x)$ ending 
in the QCP near optimal doping \cite{CAST,BECCA,ANDERGASSEN}. This proposal 
of a (dynamical) CO instability finds support from neutron scattering 
experiments in Ne-doped LSCO \cite{TRANQUADA} and ${\rm YBa_2Cu_3O_{7-x}}$ 
systems \cite{MOOK}, where charge domain walls (stripes) have been detected.

In this paper, by extending previous results, we account for the various
pseudogap-formation temperatures and give a reason why formed stripes may be 
present in some classes of cuprates and be absent in other classes. We also
address the issue of the anomalous isotopic effects of $T_c$ and $T^*$ in the 
cuprates and provide a novel mechanism based on the relevance of CO 
fluctuations.

\section{THE CHARGE-ORDERING SCENARIO AND THE CUPRATE PHASE DIAGRAM}
The above CO scenario found one of its possible realizations within a minimal 
model of strongly correlated electrons interacting with phonons, i.e., the 
one-band Hubbard-Holstein (HH) model in the presence of long-range Coulomb 
forces \cite{CAST,BECCA}. In this model there is a hopping term $t$ 
representing the electron kinetic energy, a (large) on-site Hubbard repulsion 
$U$, representing the strong electron-electron (e-e) correlation, a coupling 
$g$ between a local lattice distortion and the local electron density, a 
phonon frequency $\omega_0$, characterizing the relevant lattice distortion, 
and a long-range e-e Coulomb repulsion $V_C$. The simplification with respect 
to models taking into account the complex structure of the cuprates was 
adopted for two main reasons: first of all it makes the model manageable 
despite the formal complications related to the treatment of strong e-e 
correlations via the slave-boson large-$N$ expansion formalism. Moreover it 
allows to concentrate on the relevant aspects providing both a 
non-Fermi-liquid behavior and a strong pairing mechanism. Although the 
relevant features related to quantum criticality should be generic, different 
cuprates should of course show different aspects related to their specific 
structure. 

Two major results were obtained within the HH model \cite{CAST,BECCA}. First 
of all, at mean field, one obtains the $T=0$ phase diagram reported in Fig. 1 
with a line of the electron-phonon (e-ph) coupling $g$ as a function of the 
doping $x$. This line marks a second-order transition for the onset of CO 
characterized by an order parameter $\rho_{{\bf q}_c}$ representing the 
microscopic charge modulation at a wave vector ${\bf q}_c$ \cite{notaorder}. 
Each point on the curve $g(x)$ is a QCP, which is the $T=0$ end point of a 
critical line $T^0_{CO}(x)$. Secondly, charge collective fluctuations mediate 
a doping- and temperature-dependent singular scattering among the 
quasiparticles
\begin{equation}
\label{Gamma}
\Gamma({\bf q},\omega)  \simeq {\tilde U}-{V \over \xi_0^{-2} + 
\vert {\mathbf q}-{\mathbf q}_c \vert^2 -{\rm i}\gamma\omega},
\end{equation}
where ${\tilde U}$ is a residual repulsion, $V$ is the strength of the 
singular interaction, $\gamma \sim t^{-1}$ is a characteristic time scale of 
the charge fluctuations, and $\xi_0^{-2}$ is the inverse square correlation 
length, which measures the distance from criticality. This interaction 
diverges at $\omega=0$ and at a critical wave vector ${\bf q}_c$ when 
$\xi_0^{-2}$ vanishes. This occurs at $T=0$ when the doping is reduced down 
to a mean-field critical value $x_c^0$, and at finite $T$, when approaching 
the mean-field critical line $T^0_{CO}(x)$. As we discuss in Sec. II B, the 
CO fluctuations beyond mean field shift the mean-field critical line 
$T_{CO}^0(x)$ to a lower value $T_{CO}(x)$, ending in the 
fluctuation-corrected QCP at $x_c$ ($<x_c^0$), and modify accordingly the 
expression for the coherence length $\xi$. 

The singular part of the interaction (\ref{Gamma}) (with $\xi_0\to\xi$ if 
fluctuations are taken into account) is attractive both in the 
particle-particle and in the particle-hole channels. Therefore, the 
quasiparticles feel an increasingly strong attraction by approaching the CO 
critical line. In the particle-hole channel, this interaction can produce a 
pseudogap due to the incipient CO. At the same time in the particle-particle 
channel, the strong attraction can lead to pair formation even in the absence 
of phase coherence \cite{notacohe}. Therefore the pseudogap-formation 
temperature $T^*(x)$ closely tracks the underlying CO transition line. 

We point out that the assumption of an infinitely large Hubbard repulsion $U$
and the large-$N$ expansion do not allow for a correct description of the 
spin degrees of freedom \cite{notaaf}. However, at large but finite $U$ the 
above scenario is not expected to drastically change, except that the 
modulation of the charge profile within the CO state should favor 
antiferromagnetic ordering in the charge-poor regions. This state with charge 
and (enslaved) spin modulation mimics the stripe phase observed in the 
cuprates, when approaching it from the large-doping side, where charge 
degrees of freedom play a major role. Enslaved nearly critical 
antiferromagnetic fluctuations with a characteristic wave vector 
${\bf q}_c^{AF}\simeq(\pi,\pi)$ are expected to mediate an interaction 
similar to Eq. (\ref{Gamma}) in the particle-particle channel, which becomes 
repulsive in the Cooper channel. This additional contribution is relevant to 
properly reproduce the main features of the single-particle spectra 
\cite{LORE,SAINI}. The $d$-wave symmetry of the superconducting order 
parameter, which is already favored by the residual repulsive term in Eq. 
(\ref{Gamma}) \cite{PERALI}, is further stabilized by the antiferromagnetic
fluctuations.

%%%%%%%%%%%%%%%%%%%%%%%%%%%%%%%%%%%%%%%%%%%%%%%%%%%%%%%%%%%%%%%%%%%%%%%%%
\begin{figure}  
{\psfig{figure=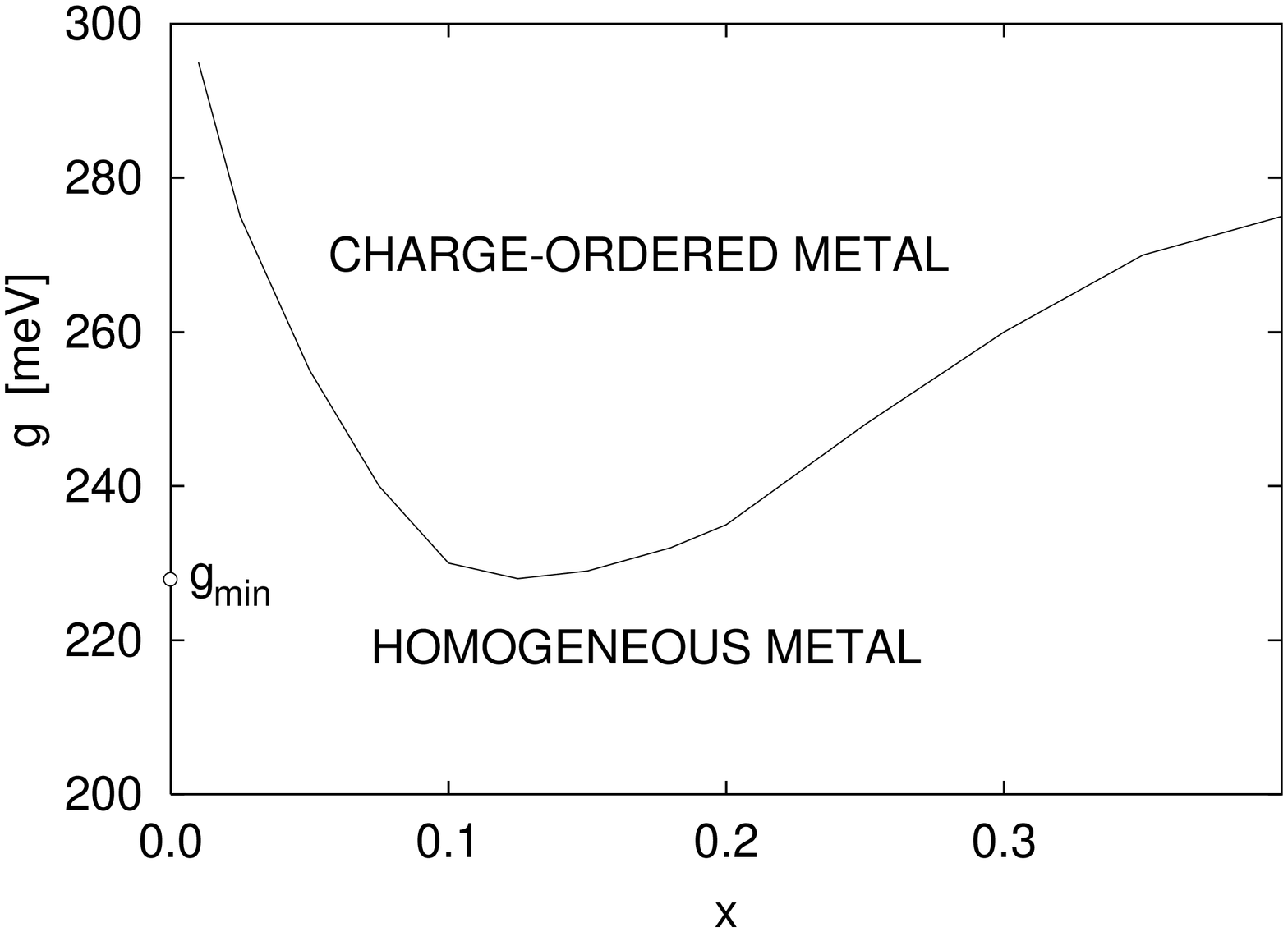,width=8.5cm,angle=-90}}  
\end{figure}
\vspace {-0.5truecm}
{\small FIG. 1. Phase diagram  e-ph coupling $g$ vs doping $x$ of the 
single-band infinite-$U$ Hubbard-Holstein model with nearest-neighbor hopping 
$t=0.5$ eV, next-to-nearest-neighbor hopping $t'=-(1/6)t$, phonon frequency
$\omega_{0}=40$ meV, and in the presence of long-range Coulomb forces with 
strength $V_{C}=0.55$ eV. See Ref. \cite{BECCA} for a detailed description of 
the model. An empty circle on the $g$ axis marks the minimum value 
$g_{\rm min}$ allowing for CO.}
%%%%%%%%%%%%%%%%%%%%%%%%%%%%%%%%%%%%%%%%%%%%%%%%%%%%%%%%%%%%%%%%%%%%%%%%%

\subsection{The zero-temperature phase diagram}
The separation between the homogeneous phase and the CO phase at $T=0$,
reported in Fig. 1, is marked by a line having a minimum value $g_{\rm min}$,
below which no CO (or stripes) could be formed as a static phase. Of course 
strong quantum fluctuations mediating the singular scattering (\ref{Gamma}),
which leads to pairing and non-Fermi-liquid behavior in the normal phase, 
would be present both above and below the quantum critical line. 

The crucial point in obtaining the CO instability is that strong electronic 
correlations reduce the effect of the homogenizing electron kinetic-energy 
term, thus favoring the possibility of an instability induced by a residual 
interaction, which in our model is provided by the e-ph interaction. The 
relevant role of the lattice in driving the instability is, indeed, strongly 
suggested by the peculiar isotopic effect of $T^*$ and $T_c$ 
\cite{ANDERGASSEN} and by recent EXAFS experiments \cite{BIANCONI}, which 
identified interesting features associated to the lattice structure. 
Specifically a microstrain $\varepsilon$ of the Cu-O bonds was measured in 
terms of the deviation of the Cu-O distance with respect to a reference 
distance $d_0$, which introduces a mismatch in the lattice between the 
CuO$_2$ layers and the rock-salt layers. Actually there is a close 
resemblance between the experimental $(\varepsilon, x)$  and the theoretical 
$(g,x)$ phase diagrams indicating that the Cu-O microstrain and the e-ph 
coupling $g$ of the simplified HH model are strictly related. From the 
microscopic point of view it is indeed quite natural that a lattice 
contraction in the CuO$_2$ planes can enhance the effective coupling between 
the electrons and the ions. Therefore the comparison between experiments and 
theory allows to draw the conclusion that a one-to-one monotonic relation 
$g=g(\varepsilon)$ is likely to exist.   

Once this general framework is settled, one can move to identify more 
detailed possible scenarios. In particular the order of the homogeneous-metal 
to charge-ordered-metal transition is of obvious relevance. The simplest 
possibility is that in the real materials this transition is of the second 
order. In this case, at $T=0$,  the different microstrains determined by the 
rock-salt layers directly correspond to different e-ph couplings and are 
reflected in the similar phase diagrams. Of course this simple theoretical
picture is not necessarily realized in the quite complex real materials  
where i) anharmonic effects and/or ii) additional non-ordering  fields can 
partially or entirely transform the second-order transition  into a 
first-order one. In this last case, the above scenario maintains its 
validity, as long as this transition is {\it weakly} first-order.

%%%%%%%%%%%%%%%%%%%%%%%%%%%%%%%%%%%%%%%%%%%%%%%%%%%%%%%%%%%%%%%%%%%%%%%%%%%
\begin{figure}
{\psfig{figure=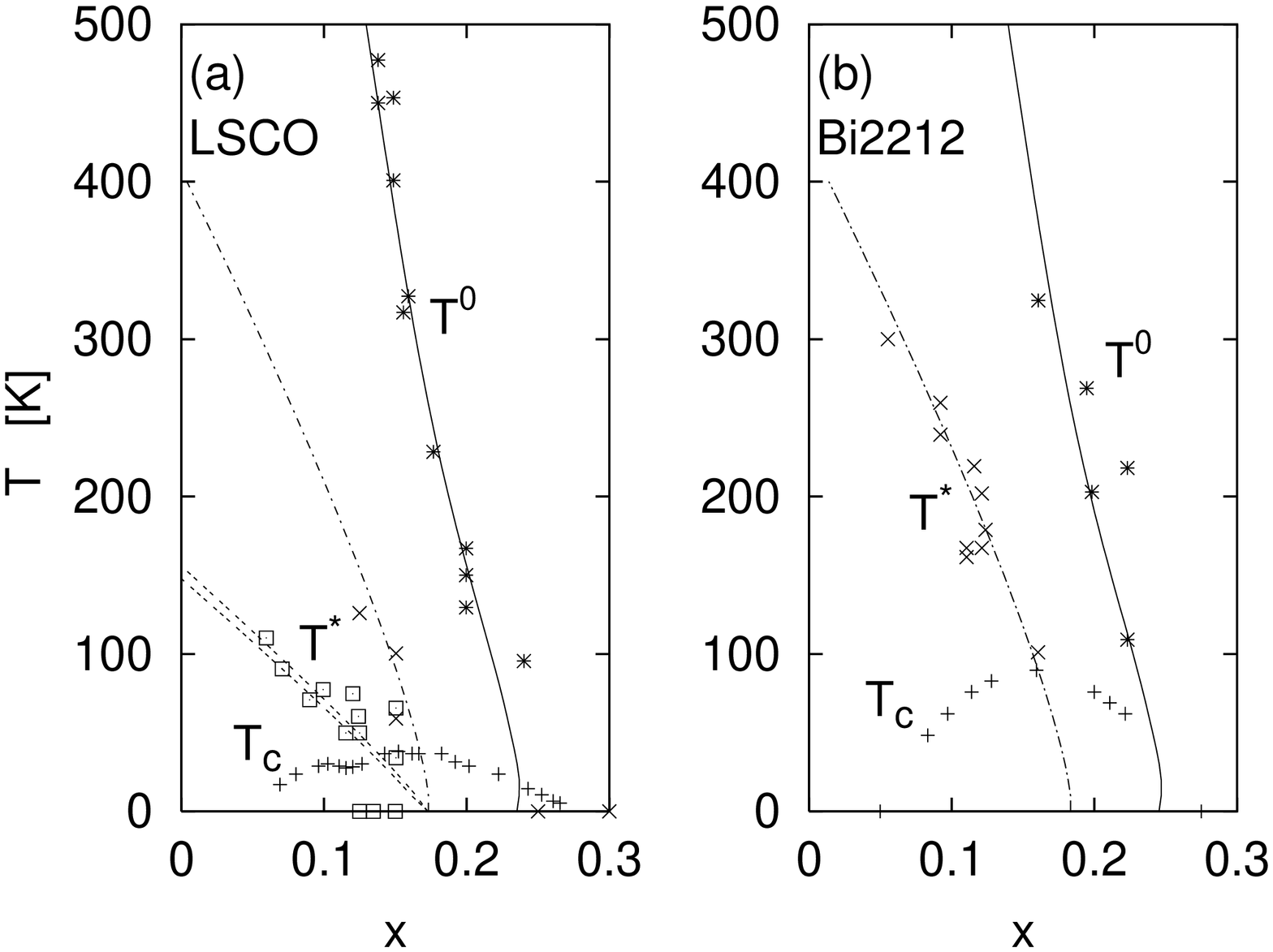,width=8.5cm,angle=-90}}
\end{figure}
\vspace {-0.5truecm}
{\small FIG. 2. 
The phase diagram of the cuprates according to the CO-QCP scenario for LSCO 
(a) and Bi2212 (b). The solid line is the m-f critical line $T_{CO}^0(x)$ 
ending at $T=0$ in the m-f QCP at $x_c^0$. The lowest dashed line in panel 
(a) marks the 3D critical line $T_{CO}(x)$ in the presence of fluctuations, 
ending in the QCP at $x_c$. The dot-dashed line in panels (a,b) indicates the 
``dynamical instability'' condition (see text) for $\omega_{probe}=1$ meV. The 
intermediate dashed line in panel (a) represents the ``dynamical 
instability'' condition for $\omega_{probe}=1\, \mu$eV. The experimental 
points for $T^0$ (stars) and for $T^*$ measured with fast (crosses) and slow 
(squares) probes for LSCO are from Ref.\cite{BILLINGE}, those for Bi2212 are 
from Refs.\cite{DING,ISHIDA}. The critical temperatures $T_c$ are also shown 
(pluses).
}
\vspace{.3cm}
%%%%%%%%%%%%%%%%%%%%%%%%%%%%%%%%%%%%%%%%%%%%%%%%%%%%%%%%%%%%%%%%%%%%%%%%%%%

We stressed in several papers \cite{RASS} that the actual onset of a fully   
developed CO phase is competing with local or coherent pair formation, which 
modifies the fermionic spectrum stabilizing the system against the electronic 
CO transition. There is the complementary possibility that the 
charge-ordered state can never be reached because $g$ is below $g_{\rm min}$. 
The corresponding observable quantity identified by the experiments in Ref. 
\cite{BIANCONI} is the critical microstrain $\varepsilon_c$, below which no 
stripe phase can be observed. Therefore the cuprates can deviate from CO 
criticality not only by $x-x_c$ in the (overdoped) quantum-disordered region, 
and by $T$ or $T-T_{CO}(x)$ in the quantum-critical or underdoped region, but 
also because the microstrain $\varepsilon$ is smaller or larger than 
$\varepsilon_c$, thereby tending to a homogeneous or an inhomogeneous phase 
respectively. As long as  $\varepsilon-\varepsilon_c$ is not too large and 
$T$ is finite, for $x\sim x_c$ the quantum-critical region, with strong 
dynamical fluctuations, can be reached in both cases. According to the 
proposed mechanism of pairing mediated by critical fluctuations, the stronger 
are these fluctuations and the larger is $T_c$. Indeed in the experiment the 
mercury compound Hg1212, having  $\varepsilon \sim \varepsilon_c$, has the 
the largest $T_c$, while Hg1201 (with $\varepsilon < \varepsilon_c$) and 
Bi2212 or LSCO (with $\varepsilon > \varepsilon_c$) have a lower $T_c$. 
Moreover, since $\varepsilon < \varepsilon_c$ in Hg1201, we expect no 
$T_{CO}(x)$, and therefore no $T^*$, for this material.

\subsection{Charge-ordering transition at finite temperature and pseudogap
formation}
In this section we calculate the temperature for the onset of the CO 
instability within the mean-field approximation, $T_{CO}^0(x)$, and its 
corrected value, $T_{CO}(x)$, in the presence of the leading fluctuations 
beyond mean field. For establishing the onset of the CO phase coming from 
high-doping and high-temperature regime, the direct involvement of the spin
degrees of freedom can be safely ignored. 

For $T<T_{CO}^{0}$ the CO fluctuations become substantial, leading to a 
reduction of the quasiparticle density of states. We therefore identify our
mean-field line $T_{CO}^0(x)$ with the weak-pseudogap crossover line $T^0(x)$ 
observed in Knight-shift, transport, and static susceptibility measurements 
\cite{BILLINGE}, as the incipient reduction of the quasiparticle density of
states. The fluctuation-corrected critical line $T_{CO}(x)$ 
[$\ll T_{CO}^0(x)$], where the correlation length $\xi$ diverges, and the 
scattering (\ref{Gamma}) becomes effectively singular, is instead closely
tracked by the pseudogap crossover line $T^*(x)$, according to the discussion
of Sec. II. 

All the complicated formal structure of the quasiparticle scattering, 
mediated by phonons and by the Coulomb interaction within the slave-boson 
approach, is, at the end, simply represented, at least near criticality, by a 
RPA resummation $\Gamma({\bf q},\omega_n)\approx V_{eff}({\bf q})/
[1+V_{eff}({\bf q})\Pi ({\bf q},\omega_n)]$ of an effective static interaction
$V_{eff}({\bf q})={\tilde U}({\bf q})+V_C({\bf q})-\lambda t$ \cite{SEIBOLD}. 
Here $\Pi$ is the fermionic polarization bubble,  
$\lambda\equiv 2g^2/t\omega_0$ is the dimensionless e-ph coupling, and 
${\tilde U}({\bf q})\simeq A+B|{\bf q}|^2$ is the residual short-range 
repulsion between quasiparticles, and $V_C({\bf q})$ is the Fourier transform
of the e-e Coulomb repulsion. For the correspondence of $V_{eff}({\bf q})$ 
with the parameters of the original HH model see Ref. \cite{SEIBOLD}. The 
mean-field CO instability condition, which occurs for reasonable values 
$\lambda \sim 1$, is $1+V_{eff}({\bf q}_c)\Pi({\bf q}_c, \omega=0)=0$. At 
$T=0$ this determines ${\bf q}_c$ and the position of the mean-field QCP 
$x_c^0$. By expanding $1+V_{eff}\Pi$ near the instability at $T=0$, we find
$\xi_0^{-2}\propto x-x_c^0$. At $T>0$, considering the $T$ dependence of 
the bare polarization bubble $\Pi$, the instability condition determines
the mean-field critical line $T_{CO}^0(x)$ (the solid-line curves in Fig. 2), 
which ends at $x_c^0$. We evaluate $T_{CO}^0(x)$ by taking standard 
quasiparticle (i.e., dressed by the slave bosons) band parameters 
($t=200$ meV, $\omega_0=70$ meV, leading to $A=200$ meV and $B=170$ meV in 
the residual repulsion $\tilde U$). $V_C=220$ meV and $g=210$ meV are 
adjusted to match with the extrapolated experimental values of $T^0(x)$ for 
LSCO \cite{BILLINGE} ($x_0\approx 0.22$) with the mean-field QCP $x_c^0$. 
Without any further adjustment of the parameters the curve $T_{CO}^0(x)$ 
agrees remarkably well with the experimental data for $T^0(x)$. Similar 
parameters are taken for Bi2212 to fit the data in Refs. \cite{DING,ISHIDA}, 
with $V_C=220$ eV and $g=230$ eV.

%%%%%%%%%%%%%%%%%%%%%%%%%%%%%%%%%%%%%%%%%%%%%%%%%%%%%%%%%%%%%%%%%%%%%%%%%%%
\begin{figure}
{\hspace{1 truecm}\psfig{figure=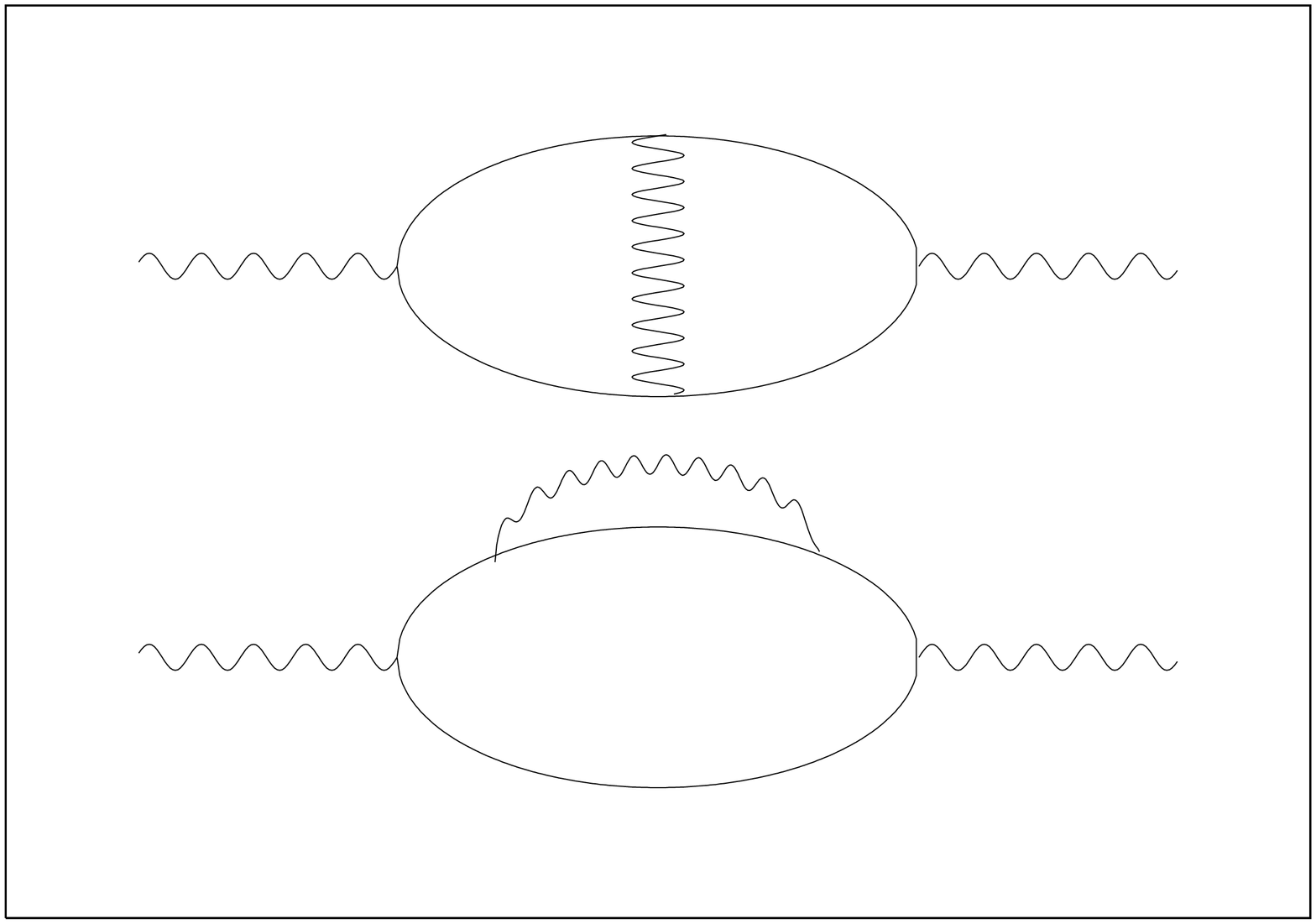,width=5.5cm,angle=-90}}
\end{figure}
\vspace {-0.5 truecm}
{\small FIG. 3. 
The vertex and selfenergy corrections beyond RPA for the fermionic bubbles 
(solid line) due to critical CO fluctuations [the wavy line represents the 
singular part of the effective interaction, Eq. (\ref{Gamma})].}
\vspace{1cm}
%%%%%%%%%%%%%%%%%%%%%%%%%%%%%%%%%%%%%%%%%%%%%%%%%%%%%%%%%%%%%%%%%%%%%%%%%%%

Due to the strong anisotropy of the layered cuprates, the contribution of the
fluctuations related to any instability is significant 
\cite{MILLIS,MARKIEWICZ}. Specifically the CO fluctuations which mediate the 
critical effective interaction, Eq. (\ref{Gamma}), can be included in the 
polarization bubble $\Pi$ (entering the instability condition) via the 
diagrams of Fig. 3, leading to corrections beyond RPA. From the explicit 
evaluation of these diagrams we find the self-consistent correction to the 
inverse square correlation length  
\begin{equation}
\xi^{-2}=\xi_0^{-2}+12{\tilde u}T\sum_{|\omega_n|<\omega_0\atop \bf q}
{V\over \xi^{-2}+|{\bf q}-{\bf q}_c|^2+\gamma|\omega_n|},
\label{mass1}
\end{equation}
where $\omega_n$ are the bosonic Matsubara frequencies, and the coupling 
constant ${\tilde u}\sim V/t^3$. Since the phonons are responsible for the 
instability, $\omega_0$ appears as the frequency cut-off \cite{BECCA}. The 
fluctuation-corrected instability condition is $\xi^{-2}=0$. As it stands, 
Eq. (\ref{mass1}) is written for the two-dimensional Cu-O planes. At $T=0$, 
Eq. (\ref{mass1}) with $\xi^{-2}=0$ leads to a finite shift of the 
two-dimensional (2D) QCP $x_c^0-x_c \propto \omega_0$ (see Fig. 2). At finite 
$T$, purely 2D fluctuations suppress the transition. However, by considering
the more realistic anisotropic 3D character of the critical fluctuations,
we obtain a finite fluctuation-corrected transition line $T_{CO}(x)$,
which is reported as the lowest dashed line in  Fig. 2(a). Thus, the 
inclusion of fluctuations brings the critical line from temperatures of the 
order of typical electronic energies ($T_{CO}^0 \sim t$) down to much 
lower temperatures $T_{CO}$ of the order of the observed $T^*$'s. 

The spread in the measured values of $T^*$ depending on the typical frequency 
of the experimental probe $\omega_{probe}$ (see, e.g., Ref. \cite{BILLINGE}) 
suggests, however, within our identification of $T_{CO}$ with $T^*$, that the 
CO instability may be ``dynamical''. In this case, the critical condition 
$\xi^{-2}=0$ is replaced by $\xi^{-2}=\gamma\omega_{probe}$ in the 
self-consistency condition, Eq. (\ref{mass1}). Two examples are reported in 
Fig.2 for $\omega_{probe}=1$ meV (dot-dashed line in Figs. 2(a,b), typical of 
neutron scattering experiments, and $\omega_{probe}=1~\mu$eV (second dashed 
line from bottom in Fig. 2(a), typical of static experiments (NQR, NMR). The 
corresponding experimental data for $T^*$ in LSCO are also reported for 
comparison. The coupling $V$ between the charge fluctuations and the 
quasiparticles, which is the only parameter for which an a priori estimate is 
difficult, is fixed by matching the position of the fluctuation-corrected QCP 
with the $T=0$ extrapolation of the experimental values of $T^*$. We used 
$\gamma=0.7$ eV$^{-1}$ and $\gamma=0.4$ eV$^{-1}$ for LSCO and Bi2212 
respectively, and $V=0.54$ eV for both. Note that, similarly to the case of 
$T_{CO}^0(x)$, the agreement between the calculated $T_{CO}(x)$ and the 
experimental points for $T^*$ is obtained without adjustment of other 
parameters. 

\section{Novel isotope effects}
In the present framework, any mechanism shifting the position of the QCP and 
the line $T_{CO}(x)$ is mirrored by a corresponding shift of the 
superconducting critical line $T_c(x)$ and of the pseudogap formation
temperature $T^*(x)$. This is the case of the shift induced by isotopic 
substitution, and we want to connect our results with the observation of 
isotopic effects on $T_c$ \cite{FRANCK,CRAWFORD} and on $T^*$ 
\cite{RAFFA,WILLIAMS,RUBIO}. As shown in Sec. II A, the mean-field 
weak-pseudogap crossover temperature $T^0_{CO}\sim T^0$ depends on the phonon 
frequency only through the dimensionless e-ph coupling $\lambda$, which
is known to display no isotopic dependence. Therefore $T^0_{CO}$, and
consequently $T^0$ and the portion of $T_c(x)$ near and above $T^0(x)$, are 
not expected to display isotope effects. 

On the other hand, the CO fluctuations, as shown above, crucially involve 
$\omega_0$, and new physical effects can be induced by the isotopic 
substitutions, on fluctuation-dependent physical quantities, such as $x_c$ 
and $T_{CO}(x)$. Since $\omega_0$ decreases with increasing ionic mass, 
the effect of fluctuations is reduced, and $x_c$ and $T_{CO}(x)$ remain 
closer to their mean-field values. In Fig. 4, we report the line $T_{CO}(x)$, 
calculated via Eq. (\ref{mass1}), with the same parameters used in Sec. II B 
to fit the $T^*$ data of LSCO, together with its isotopic shift calculated 
for $^{16}O\to \, ^{18}O$ substitution (i.e., for a five percent reduction of 
$\omega_0$). 

%%%%%%%%%%%%%%%%%%%%%%%%%%%%%%%%%%%%%%%%%%%%%%%%%%%%%%%%%%%%%%%%%%%%%%%%%%%
\begin{figure}
{\hspace{0.5 truecm}{\psfig{figure=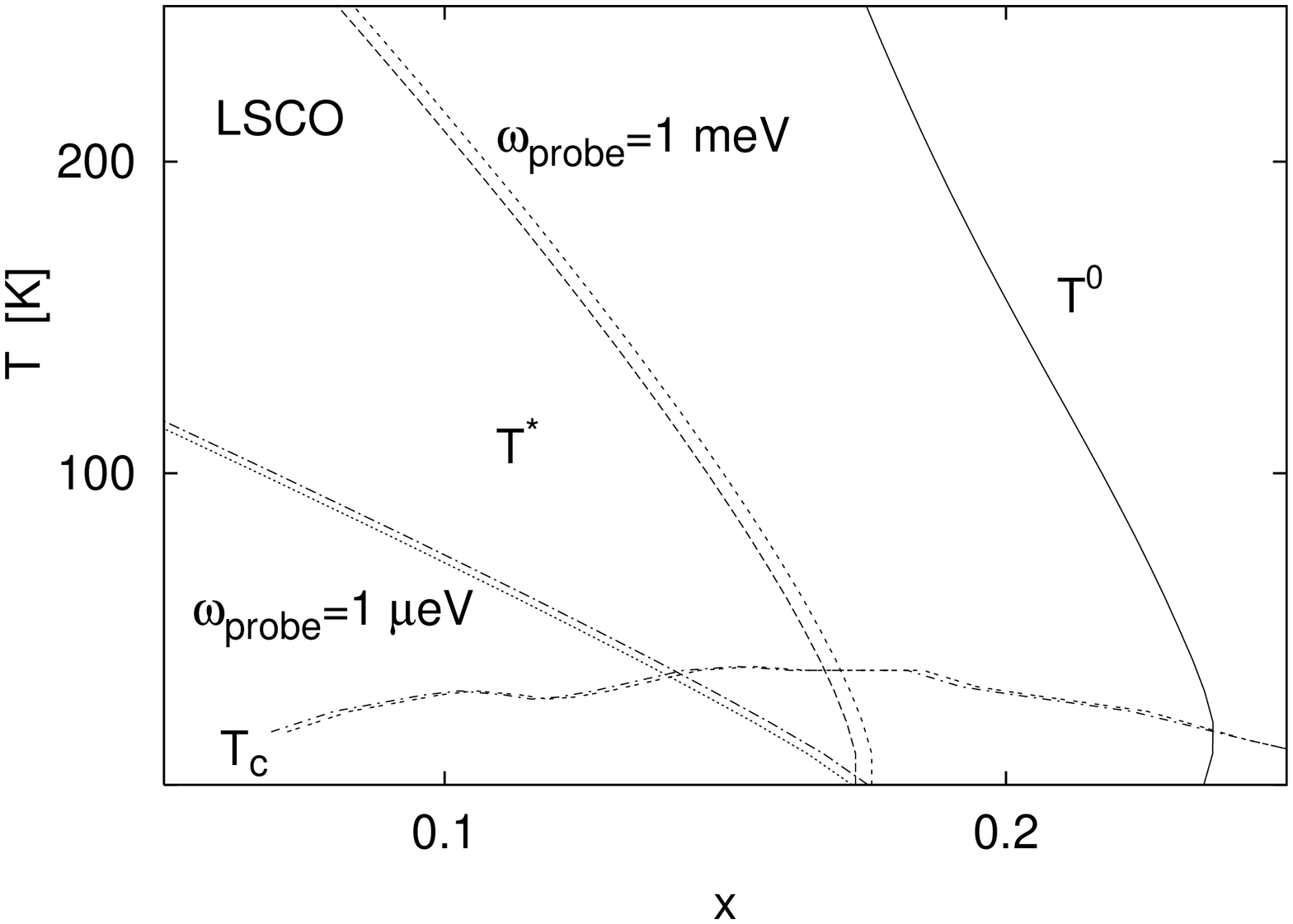,width=7.cm,angle=-90}}}
\end{figure}
{\small FIG. 4. 
Calculated effect of the isotopic change ${\rm ^{16}O \to ^{18}O}$ 
(i.e. $\omega_0=70$ meV and $\omega_0'=66$ meV) on $T^*$ in LSCO, both for a 
fast- and a slow-probe measurement. The inferred shift on $T_c$ is also 
reported, although hardly visible on this scale.}
\vspace{.5cm}
%%%%%%%%%%%%%%%%%%%%%%%%%%%%%%%%%%%%%%%%%%%%%%%%%%%%%%%%%%%%%%%%%%%%%%%%%%%

Contrary to standard theories based on CO pseudogap \cite{EREMIN}, the 
isotopic shift is positive (i.e. anomalous) for $T_{CO}\sim T^*$. Moreover, 
if the slope of $T_{CO}(x)$ is large, a small isotopic shift in $x_c$ can 
result in a substantial shift in $T_{CO}\sim T^*$, i.e., the steeper is 
$T^*$, the larger is the isotopic effect. Since the slope of $T_{CO}(x)$ 
increases by increasing $\omega_{probe}$, whereas the QCP is unshifted, the 
isotope effect on $T_{CO}\sim T^*$ is enhanced, and we have the general trend 
that faster probes should detect a larger isotope effect on $T^*$. Although 
we cannot account for the near-absent or negative isotope effect on $T^*$ 
within the almost static probes in ${\rm YBa_2Cu_4O_8}$ 
\cite{RAFFA,WILLIAMS}, this general trend is in qualitative agreement with a 
much stronger effect observed in the isostructural ${\rm HoBa_2Cu_4O_8}$
(HBCO-124) with fast neutron scattering \cite{RUBIO}. Indeed, this fast-probe 
experiment should be represented by the curve $T_{CO}(x)$ corresponding to 
$\omega_{probe}=1$ meV in Fig. 4. The huge isotope effect on $T^*$ observed 
in HBCO-124 suggests that the curve $T_{CO}(x)$ in this system is steeper 
than in LSCO.

As far as $T_c$ in the underdoped region is concerned, we can only infer some 
expected consequences. Assuming that the shift in $x_c$ produces a rigid 
shift of $T_c(x)$ along the doping axis, the isotopic effect for $T_c$ is 
negative (i.e. normal) upon reducing $\omega_0$, and much smaller than for 
$T^*$ (the $T_c(x)$ curve being much flatter), in agreement with 
long-standing experiments \cite{FRANCK,CRAWFORD}. This large difference in 
the isotope effect for $T_c$ and $T^*$, $\left(\Delta T_c/\Delta M\right)/
\left(\Delta T^*/\Delta M\right)\ll 1$, is indeed experimentally observed in  
HBCO-124, and reported in Ref. \cite{RUBIO}, where  a ``striking similarity 
between isotopic substitution and underdoping with respect to both $T_c$ and 
$T^*$'' is pointed out. Although we are not aware of any systematic analysis 
of the doping dependencies of $T_c$, $T^*$ and their isotopic shifts in 
HBCO-124, allowing for a strict comparison, this observation finds its natural 
interpretation within our theory, where the isotopic substitution 
increases $x_c$ and is therefore nearly equivalent to underdoping.

In the strongly underdoped materials, where substantial isotope effects in 
the penetration depth \cite{HOFER} and in the x-ray absorption \cite{LANZARA} 
have been observed, our results can provide only qualitative indications, as 
other effects (magnetic, polaronic, lattice-pinning) not included in 
our HH model become relevant. 

%%%%%%%%%%%%%%%%%%%%%%%%%%%%%%%%%%%%%%%%%%

\section{Conclusions}

In conclusion we have revisited the CO scenario, and explicitly evaluated
the mean-field and the fluctuation-corrected critical lines, $T_{CO}^0(x)$
and $T_{CO}(x)$, for the onset of the charge inhomogeneous phase, which 
should represent the stripe phase of the cuprates, when present. The nearly 
singular effective interaction among the electrons mediated by the CO 
fluctuations near the instability, suggests the natural identification of the 
experimentally observed weak-pseudogap crossover line $T^0(x)$ with 
$T_{CO}^0(x)$, and of the pseudogap formation temperature $T^*$ with 
$T_{CO}(x)$. We obtain in this way a good fitting of the experimental data, 
and by assuming that the CO transition has a dynamical character, we account 
for the spread of the measurements of $T^*(x)$, depending on the 
characteristic time scale of the experimental probe. The existence of the CO 
transition, depending on the strength of the e-ph coupling, as produced by 
our zero-temperature phase diagram, can give an answer to the puzzling 
question why features of formed stripes are present in some cuprates and 
absent in others. The CO instability condition at mean-field level depends 
only on the dimensionless e-ph coupling which is not changed by isotopic 
substitution. Therefore, in our model, $T_{CO}^0(x)$ should not undergo any
isotopic shift. On the contrary, $T_{CO}(x)$, and therefore $T^*$, are strongly
modified by the effect of CO fluctuations, which in our approach is 
proportional to the phonon frequency $\omega_0$. Within our model, therefore,
$T^*$ shows a novel isotope effect produced by charge fluctuations, with a 
positive shift, which depends on the slope of $T_{CO}(x)$. It results, 
therefore, that the isotope effect is stronger when detected with faster 
probes. We also infer that $T_c(x)$ should have no isotope effect in the 
optimal and overdoped regimes, and a small negative effect in the underdoped
region, which increases upon underdoping. In this new isotope effect, the 
phonon do not appear directly as mediators of pairing, but are indirectly 
involved, via the CO fluctuations, and qualitatively explain the complex 
behavior of $T^*$ and $T_c$ upon isotopic substitution. 
\vskip 0.5truecm
  
{\em Acknowledgments.} 
We acknowledge stimulating discussions with C. Castellani.

%\vspace {-0.5 truecm}
%%%%%%%%%%%%%%%%%%%%%%%%%%%%%%%%%%%%%%%%%%%%%%%%%%%%%%%%%%%%%%%%%%%%%%
   
\end{multicols}  

\begin{references}
\bibitem{notaorder}In the framework of Refs. \cite{CAST,BECCA} the CO
        instability is a second-order transition leading to a gradual 
        increase of the charge modulation. It is only when one enters deeply 
        inside the (locally) ordered phase that anharmonic distortions of the 
        charge profile may arise from the enhanced interactions with the spin 
        and the lattice degrees of freedom. This may lead to the stripe 
        formation.
\bibitem{notaaf} For the same reason, the low-doping region in Fig. 1 is only 
        indicative.
\bibitem{notaz2} This feature is the natural consequence of the damping of 
        the charge fluctuations due to the creation of particle-hole pairs at 
        arbitrarily low energy when ${\mathbf q}_c$ joins two ``hot'' points 
        on the Fermi surface.
\end{references}
\end{document}